\documentclass[12pt,preprint]{aastex}
\newcommand{\HI}{H~{\sc i}} 

\newcommand{\kms}{${\rm km~s^{-1}}$}

\slugcomment{{\em To appear in Astrophysical Journal Letters}}
\shortauthors{McCLURE-GRIFFITHS ET AL.} 
\shorttitle{}

\begin{document} 

\title{An Interaction of a Magellanic Leading Arm High Velocity Cloud with the Milky
  Way Disk}

\author{N.\ M.\ McClure-Griffiths,\altaffilmark{1} L.\
  Staveley-Smith,\altaffilmark{2} Felix  J.\ Lockman,\altaffilmark{3} M.\
  R.\ Calabretta,\altaffilmark{1} H.\ Alyson Ford,\altaffilmark{1,4} P.\ M.\
  W.\ Kalberla,\altaffilmark{5} T. Murphy,\altaffilmark{6,7}
  H. Nakanishi,\altaffilmark{1,8} D.\ J.\ Pisano\altaffilmark{3} }

\altaffiltext{1}{Australia Telescope National Facility, CSIRO,
  Marsfield NSW 2122, Australia; naomi.mcclure-griffiths@csiro.au,
  mark.calabretta@csiro.au}

\altaffiltext{2}{School of Physics, University of Western Australia,
  Crawley WA 6009, Australia; lister.staveley-smith@uwa.edu.au}

\altaffiltext{3}{National Radio Astronomy Observatory, Green
  Bank, WV 24944; jlockman@nrao.edu, dpisano@nrao.edu}

\altaffiltext{4}{Centre for Astrophysics and Supercomputing, Swinburne
  University of Technology, Hawthorn VIC 3122, Australia;
  alyson@astro.swin.edu.au}

\altaffiltext{5}{Argelander-Institut f\"ur Astronomie, Universit\"at
  Bonn, 53121 Bonn, Germany; pkalberla@astro.uni-bonn.de}

\altaffiltext{6}{School of Physics, University of Sydney, NSW 2006,
  Australia; tara@physics.usyd.edu.au}

\altaffiltext{7}{School of Information Technologies, University of
  Sydney, NSW 2006, Australia}

\altaffiltext{8}{Present address: Faculty of Science, Kagoshima
  University, Kagoshima 890-0068, Japan; hnakanis@sci.kagoshima-u.ac.jp}

\begin{abstract}
  The Leading Arm of the Magellanic System is a tidally formed \HI\
  feature extending $\sim 60\arcdeg$ from the Magellanic Clouds ahead
  of their direction of motion. Using atomic hydrogen (\HI) data from
  the Galactic All Sky-Survey (GASS), supplemented with data from the
  Australia Telescope Compact Array, we have found evidence for an
  interaction between a cloud in the Leading Arm and the Galactic disk
  where the Leading Arm crosses the Galactic plane.  The interaction
  occurs at velocities permitted by Galactic rotation, which allows us
  to derive a kinematic distance to the cloud of 21 kpc, suggesting
  that the Leading Arm crosses the Galactic Plane at a Galactic radius
  of $R\approx 17$ kpc.  
\end{abstract}

\section{Introduction}
\label{sec:intro}
The Magellanic System is engaged in a complicated interaction with the
Milky Way, which has created a trailing ``Magellanic Stream''
\citep{wannier72}, and a tidal ``Leading Arm'' of \HI.  The Leading
Arm has a metallicity similar to that of the Magellanic Clouds and a
consistent velocity structure from its origin in the Clouds at $b
\lesssim -25\arcdeg$, through the Galactic equator, to $ b \approx
+30\arcdeg$ \citep{mathewson74, putman98, lu98, bruens05}.  Model
orbits for the Magellanic Clouds \citep{yoshizawa03, connors06} have
predicted that the Leading Arm should cross the Milky Way disk some 30
- 60 kpc from the Sun, but there are no observational constraints on
this distance.  New proper motion measurements suggest that the
Magellanic Clouds may be moving faster than previously thought,
implying a larger crossing distance for the Leading Arm than earlier
models \citep{kallivayalil06a,kallivayalil06b}.

High velocity cloud HVC 306-2+230, centered at $(l,b,v)=(305\fdg7,\,\,
-0\fdg8, \, \, 220$ \kms), is one of the many clouds that make up the
Magellanic Leading Arm \citep{mathewson79,bruens05} and is of
particular interest because it is one of two that cross the Galactic
equator.  \citet{bruens05} suggested that this HVC shows evidence of a
ram-pressure interaction with the Galactic disk.  Using new \HI\ data
from the Galactic All-Sky Survey (McClure-Griffiths et al. 2006;
McClure-Griffiths et al. 2007, in prep)\nocite{mcgriff06a}, together
with high resolution \HI\ data from the Australia Telescope Compact
Array (ATCA), we present evidence of interaction between HVC 306-2+230
and the disk of the Milky Way (\S \ref{sec:hvcimpact}).  We use the
interaction to derive a kinematic distance to the Leading Arm where it
crosses the Galactic Plane (\S \ref{subsec:dist}).  Finally, in \S
\ref{sec:orbit} we discuss the implications the distance places on the
orbit of the Magellanic Clouds.
\section{Data}
\label{sec:data}
The \HI\ data discussed here were obtained as part of the Galactic
All-Sky Survey (GASS; McClure-Griffiths et al. 2006; McClure-Griffiths
et al. 2007, in prep), which used the 21cm multibeam receiver on the
Parkes Radio Telescope to map Galactic and Magellanic \HI\ emission
over the entire sky south of Declination $\delta=0\arcdeg$.  The
project will be described in full detail in a separate paper.  For the
purposes of the current paper, the techniques for data-taking and
analysis are the same as in \citet{mcgriff06a}.  Since the 2006 paper,
the full dataset has been obtained and is used for this analysis, with
an angular resolution of 15\arcmin, channel spacing of $0.8$ \kms, and
an rms brightness temperature 1-$\sigma$ noise of $\sim 60$ mK.

We also make use of high resolution \HI\ data from the Australia
Telescope Compact Array (ATCA).  We have combined archival data from
the Southern Galactic Plane Survey \citep[SGPS;][]{mcgriff05} together
with a new 10 hr ATCA observation with the hybrid array configuration,
H168.  The SGPS data consist of observations obtained with the 375,
750A, B, C, and D array configurations of the ATCA.  The
multi-pointing ATCA data were jointly imaged using natural weighting,
deconvolved, and restored with a synthesised beam of $100\arcsec$, and
combined with the low resolution data from GASS for sensitivity to
angular scales from $100\arcsec$ to several degrees.  The ATCA data
are limited to Galactic latitudes $|b|<1.0\arcdeg$ and have an rms
brightness temperature 1-$\sigma$ noise of $\sim 1.3$ K.

\section{Evidence of a High Velocity Cloud Impact with the Galactic Disk}
\label{sec:hvcimpact}
\begin{figure}

\centering
\includegraphics[width=6.5in]{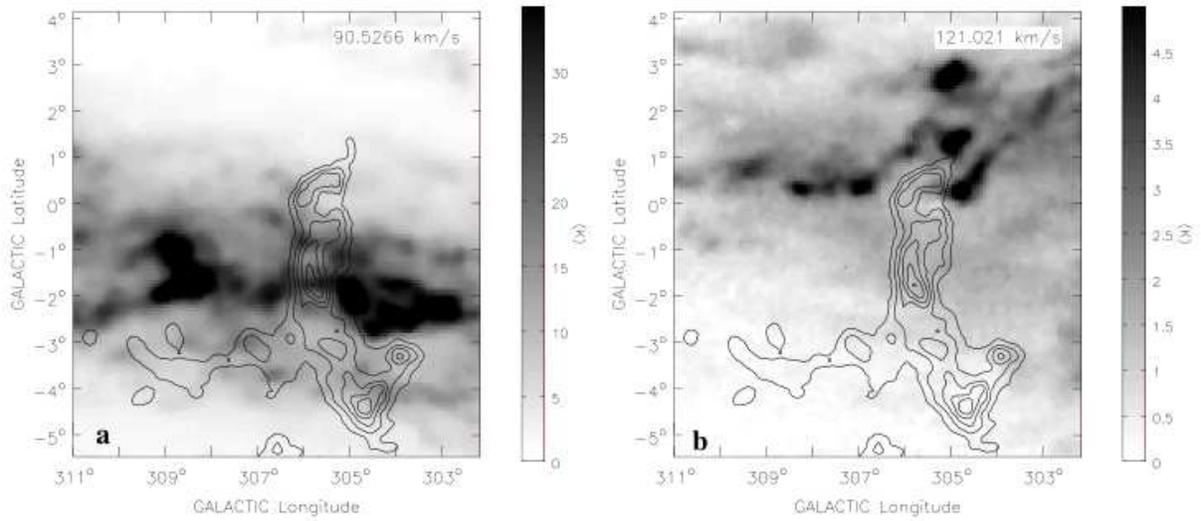}
\caption[]{ \HI\ emission from the Galactic disk (greyscale) overlaid with
  column density, $N_H$, contours of HVC 306-2+230 showing evidence for the interaction between the HVC and the Galactic disk.   ({\em a}) \HI\
  emission at $v=91$ \kms\ and ({\em b}) $v=122$ \kms.  The HVC
  includes all emission over  the velocity range
  $200 - 324$ \kms\ and contours are  displayed from $2-14\times10^{19}~{\rm cm^{-2}}$ in intervals of  $2\times10^{19}~{\rm cm^{-2}}$.  The greyscales are linear with the ranges shown in the wedges at the right of each panel.  
  \label{fig:LA_lb}}
\end{figure}

\begin{figure}
\centering
  \includegraphics[angle=-90,width=4in]{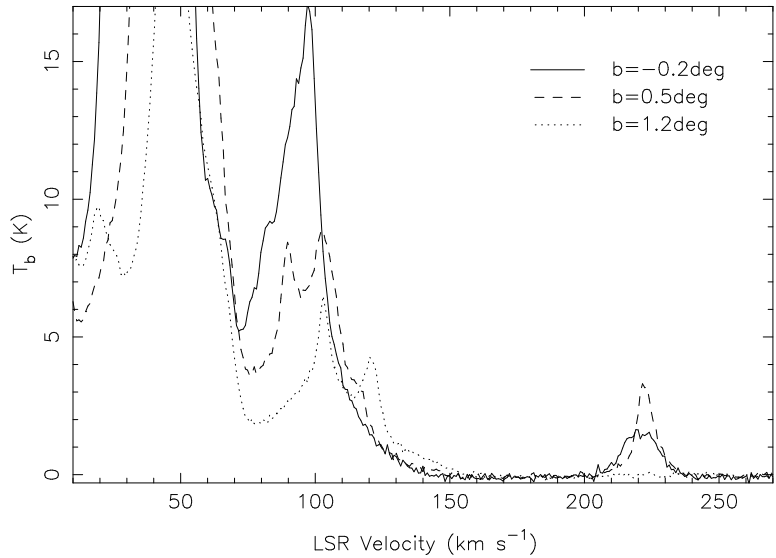}
  \caption[]{\HI\ spectra at $l=305.8\arcdeg$ towards $b=-0.2\arcdeg$ (solid line),
  $b=0.5\arcdeg$ (dashed line), and $b=1.2\arcdeg$ (dotted line),
  through the  HVC tail, the HVC head and the ridge off the dense end
  of the HVC, respectively.  All
  spectra show a peak at $v_{LSR}\sim 100$ \kms, but the spectra
  towards the cap and the head of the HVC show an additional feature
  near $v_{LSR}\sim 120$ \kms, which seems to have a spatial
  correlation with the HVC and which we interpret as arising from Galactic
  disk gass swept up by the HVC.
  \label{fig:spec}}
\end{figure}

We have used \HI\ data from GASS to search for any indication of
interaction between HVC 306-2+230 and the Milky Way disk.  Figure
\ref{fig:LA_lb} shows Galactic \HI\ emission at $v_{LSR}=91$ \kms\ and
$v_{LSR}=122$ \kms, overlaid with the column density contours of HVC
306-2+230.  There is a morphological match between the Galactic \HI\
emission and the HVC.  At $v_{LSR}=91$ \kms\ the Galactic \HI\ traces
the bottom and sides of the HVC, while at $v_{LSR}=122$ \kms\ a ridge
of Galactic \HI\ wraps around the head of the cloud.  The high
resolution data show that the ridge is very thin, only $3 -5$
arcminutes, and seems to be cold.  Spectra taken through the ridge
($b=1.2\arcdeg)$, the head of the HVC ($b=0.5\arcdeg$) and its tail
($b=-0.2\arcdeg$) are shown in Figure \ref{fig:spec}.  All show a
strong, broad feature at $v\approx 100$ \kms, but towards the ridge
there is an additional narrow feature at $v=122$ \kms. The \HI\
linewidth in the ridge is $\Delta v \sim 4$ \kms\ (FWHM), implying a
temperature of $T\leq350$ K.  There is evidence for the development of
the 122 \kms\ feature near the head of the HVC at a latitude of
$b=0.5\arcdeg$.

The context of the interaction is shown in Figure 3, in a
longitude-velocity cut through the data at $b=0.37\arcdeg$, near the tip
of the cloud.  The \HI\ around 120 \kms, which forms the ridge in
Figure 1({\em b}), is seen to be confined to longitudes between HVC
306-2+230 and a second Leading Arm cloud at $l=312\arcdeg$.  The data
thus suggest that the HVCs are interacting with gas originally at
$v_{LSR} \approx 90$ \kms\ to produce the ridge near $v_{LSR}=122$
\kms\ around the head of the HVC.

HVC 306-2+230 itself shows evidence of interaction.  The
high-resolution ATCA data of Figure~\ref{fig:LA_sgps} show that it has
a very bright, $T_b =15-20$ K, narrow bow-shaped ridge of emission
along the head of the cloud with dense clumps along the bent tail.
The column density along the bright ridge is N$_{\rm HI}\sim 10^{20}$
cm$^{-2}$. Spectra toward the head of the cloud have a two
velocity-component structure; one very narrow component of FWHM
$\Delta v=4.9 $ \kms\ and a second, broader component of width $\Delta
v= 16.9$ \kms.  This is consistent with a two-phase structure of the
cloud, where one component has $T\leq 600$ K and the other $T\leq
6000$ K. The velocity width of the spectra increases with distance
along the tail; spectra at $b=-1.6\arcdeg$ and $b=-2.4\arcdeg$ have
FWHM of $\Delta v=21.6$, and $27.8$ \kms, respectively.

The structure of the disk emission and the HVC resemble simulations of
HVCs impacting the Galactic disk which predict that a moderately dense
HVC travelling at $\sim 100$ \kms\ with respect to the disk, will form
a hemispherical shell of cool, dense \HI\ ahead of it
\citep{tenorio87, comeron92}.  The newly formed shell will be
accelerated by the interaction.  High resolution simulations of HVCs
moving through uniform media show that a very thin, bow-shaped
structure develops along the leading edge of the cloud
\citep{quilis01,konz02,agertz07}, much like the high resolution
structure of the HVC shown in Figure \ref{fig:LA_sgps}.  HVC disk
impact simulations also predict that there should be a layer of hot,
ionized gas between the neutral HVC and the neutral disk.  Although
ionized gas around HVC 306-2+230 would not be detectable in H$\alpha$
or soft X-ray emission because of absorption by foreground gas in the
Galactic plane, there is a significant offset between the head of the
HVC and the Galactic \HI\ at $v_{LSR}= 122$ \kms, which is consistent
with the presence of a layer of ionized gas.  The separation between
the HVC head and \HI\ ridge in the high resolution ATCA data is $\sim
0.6\arcdeg$, or $\sim 200$ pc at $d=21$ kpc.  The agreement between
the observations and simulations of HVC cloud impacts is additional
evidence that HVC 306-2+230 is interacting with the Milky Way disk.
\subsection{Distance to the HVC}
\label{subsec:dist}
The data presented here suggest that the HVC interacts with gas
initially at $v_{LSR} \approx 90$ \kms, which implies that the
interaction takes place well into the outer Galaxy.  We can use the
velocity of the interaction region to estimate a kinematic distance to
the HVC.  Because the HVCs are at a more positive velocity than disk
gas, any transfer of momentum from the HVC to the disk will result in
a more positive velocity for the disk gas, erroneously increasing the
kinematic distance.  Our distance estimates are thus upper limits.
Assuming a flat rotation curve, where $\Theta(R)=\Theta_0=220$ \kms\
and a Sun-Galactic Center distance of $R_0=8.5$ kpc, we estimate that
the emission is at a Galactocentric radius of $R\approx17$ kpc and a
heliocentric distance of $d\approx 21$ kpc. Use of a different
rotation curve or Sun-center distance (e.g., 8.0 kpc) changes these
values by no more than 20\%.
 
That we observe neutral gas with which the HVC is interacting also
provides an upper limit to the distance by placing it within the
detectable \HI\ disk of the Galaxy.  Although the exact extent of the
Galactic \HI\ disk is uncertain, estimates are that by 50 kpc radius
the \HI\ surface density is less than $10^{-2}~{\rm
  M_{\odot}~pc^{-2}}$ \citep{knapp78} and the majority of the \HI\ is
within 30 kpc \citep{nakanishi03,levine06b}.  The ridge of \HI\ at 120
\kms, if interpreted as material swept-up by the HVC, can give some
information on the environment the HVC is encountering.  The ridge has
$N_{\rm HI} = 7- 10 \times 10^{19}$ cm$^{-2}$ so if its line-of-sight
extent is similar to its transverse extent ($\approx 60\arcmin$) then
the column density in the $z$-direction ($< 15\arcmin$, perpendicular
to the motion of the HVC) is independent of distance and simply the
observed N$_{\rm HI}$ scaled by the ratio of widths, $15/60$.  Given
the maximum vertical extent of the ridge, the column density
perpendicular to the line of sight is $N_H< 2 - 4 \times 10^{19}$
cm$^{-2}$, equivalent to $1-3 \times 10^{-1}~{\rm M_{\odot}~pc^{-2}}$.
This is an incomplete accounting of the disk gas the HVC would
encounter, neglecting inhomogeneities in the medium and the mass of
ionized and molecular material, but nonetheless  indicates that
the upper limit on the HVC distance is $R<30$ kpc.

\subsection{HVC Drag Interaction}
Figure~\ref{fig:LA_bv} is a latitude-velocity slice taken at
$l=305.6\arcdeg$ through the center of HVC 306-2+230.  The cloud shows a
velocity gradient of about -30 \kms\ between $b=-4.4\arcdeg$, where
the central velocity is $v_{LSR}\approx252$ \kms, and $b=-1.5\arcdeg$,
where $v_{LSR}\approx223$ \kms.  The gradient is large and in the
wrong sense to arise from either the projection effects of the Sun's
motion or the upward (toward positive latitudes) motion of the HVC as
implied by its head-tail morphology and the expected trajectory of the
Leading Arm.  The gradient must be intrinsic to the HVC, and may arise
from a deceleration as the HVC goes through the disk of the Milky Way.
\citet{benjamin97} showed that drag forces should slow a cloud as it
nears the Galactic plane. We cannot measure the $z$ velocity of the
HVC, but we can estimate the change in velocity with height if the
cloud were falling at the terminal velocity; given in equation 2 of
\citet{benjamin97}.  For a cloud of column density $\sim
2\times10^{19}~{\rm cm^{-2}}$, approaching a Gaussian disk with a
midplane density $n=0.04~{\rm cm^{-3}}$ and a scale height of 1 kpc,
the difference in terminal velocity between $z=1400$ pc and $z=200$ pc
is $\sim 30$ \kms, which matches the observed velocity variation along
the length of the cloud.  This adds further support to the evidence
that the HVC is interacting with the Galactic disk; at very large
Galactic radii there would be insufficient gas to cause significant
drag.

\section{Discussion of the Magellanic Orbits}
\label{sec:orbit}
Many simulations of the Magellanic System have attributed the origin
of the Leading Arm to a close pass between the SMC and the Milky Way
which stripped gas from the SMC that was eventually pulled into a
leading tidal feature \citep[e.g.][]{connors06,yoshizawa03}.
Alternative theories give the Leading Arm an LMC origin
\citep[e.g.][]{mastropietro05}.  All of these simulations, however,
rely on a past close encounter between the Magellanic Clouds and the
Milky Way.  From this encounter the \citet{connors06} and
\citet{yoshizawa03} simulations have predicted that the Leading Arm
crosses the Galactic plane at a distance of $30 - 60$ kpc from the
Sun. Our newly derived distance of 21 kpc is smaller than the lower
limit of these predictions.

These models do not take into account recently revised measurements of
the proper motions of the LMC and SMC, which have shown that the
Magellanic Clouds have much higher tangential velocities than previous
thought \citep{vandermarel02,kallivayalil06a}.  Re-calculation of the
LMC orbit based on these proper motions and new mass models for the
Milky Way suggest that the LMC may be on its first pass and currently
at perigalacticon \citep{besla07}.  If true, and the SMC has a similar
orbit, this would have profound implications on the formation models
for the Leading Arm.  Rather than relying on close past interactions
with the Milky Way to strip LMC/SMC gas, these orbits would require an
alternative method of extracting gas from the galaxies, such as the
blowout model suggested by \citet{nidever07}.

\citet{besla07} do not predict the future orbit of the LMC, but we can
make a very simple estimate of where the LMC will cross the Galactic
plane by extrapolating from the current position, assuming a constant
LMC space velocity and neglecting Milky Way gravity.  In a Cartesian
system centred on the Galactic center where the LMC is at
$(x_0,y_0,z_0)=(-1,-41,-27)~{\rm kpc}$, with a velocity vector
$(v_x,v_y,v_z)=(-86,-268,+252)$ \kms\ \citep{kallivayalil06a},
assuming that the time to reach the Galactic plane is $t=-z_0/v_z$,
its position at $t$ will be $(-10,-70,0)$ kpc or $R \sim 71$
kpc. Because we have excluded the gravitational effect of the Milky
Way this estimate will be an upper limit.  Finally, it has been shown
that tidal features can deviate significantly from the orbits of their
parent satellites \citep{connors06,choi07}, which may also help
reconcile the new proper motions with our Leading Arm crossing
distance.
\section{Summary}
\label{sec:summary}
We have found evidence for an interaction between one of the Leading
Arm High Velocity Clouds, HVC 306-2+230, and the Milky Way disk. As
the Leading Arm crosses the Galactic equator it produces a ridge of
\HI\ at a velocity of $v_{LSR}=122$ \kms.  Further evidence for the
interaction is given by the head-tail structure of the HVC itself, its
varying velocity width, the deceleration of the cloud head, and the
agreement between simulations of HVC impacts and the morphology of the
HVC and the disk gas.  As a rare example of an HVC caught interacting
with the disk, HVC 306-2+230 offers a unique opportunity to compare in
detail observations with simulations of HVC impacts; this will be the
topic of a future paper.  The interaction with \HI\ at Galactic
velocities has allowed us to estimate a kinematic distance to the
Leading Arm of $d\approx21$ kpc from the Sun or a Galactocentric
radius of $R\approx17$ kpc, with an upper limit of $R<30$ kpc.  This
distance is close to that predicted by the \citet{connors06} and
\citet{yoshizawa03} models of the Magellanic system, and smaller than
might be expected from the \citet{kallivayalil06a} LMC proper motion
measurements.  Our new Leading Arm distance will provide an important
constraint to future models of the Magellanic System as they take into
account revised LMC and SMC proper motions.

\acknowledgements The Parkes Radio Telescope and the Australia
Telescope Compact Array are part of the Australia Telescope which is
funded by the Commonwealth of Australia for operation as a National
Facility managed by CSIRO.  We are grateful to P.\ Edwards for
the allocation of Director's Time at the ATCA to image this HVC and to
an anonymous referee for helpful comments.

\bibliographystyle{apj} 


\begin{thebibliography}{26}
\expandafter\ifx\csname natexlab\endcsname\relax\def\natexlab#1{#1}\fi

\bibitem[{{Agertz} {et~al.}(2007){Agertz}, {Moore}, {Stadel}, {Potter},
  {Miniati}, {Read}, {Mayer}, {Gawryszczak}, {Kravtsov}, {Nordlund}, {Pearce},
  {Quilis}, {Rudd}, {Springel}, {Stone}, {Tasker}, {Teyssier}, {Wadsley}, \&
  {Walder}}]{agertz07}
{Agertz}, O., {Moore}, B., {Stadel}, J., {Potter}, D., {Miniati}, F., {Read},
  J., {Mayer}, L., {Gawryszczak}, A., {Kravtsov}, A., {Nordlund}, {\AA}.,
  {Pearce}, F., {Quilis}, V., {Rudd}, D., {Springel}, V., {Stone}, J.,
  {Tasker}, E., {Teyssier}, R., {Wadsley}, J., \& {Walder}, R. 2007, \mnras,
  380, 963

\bibitem[{{Benjamin} \& {Danly}(1997)}]{benjamin97}
{Benjamin}, R.~A. \& {Danly}, L. 1997, \apj, 481, 764

\bibitem[{{Besla} {et~al.}(2007){Besla}, {Kallivayalil}, {Hernquist},
  {Robertson}, {Cox}, {van der Marel}, \& {Alcock}}]{besla07}
{Besla}, G., {Kallivayalil}, N., {Hernquist}, L., {Robertson}, B., {Cox},
  T.~J., {van der Marel}, R.~P., \& {Alcock}, C. 2007, \apj, 668, 949

\bibitem[{{Br{\"u}ns} {et~al.}(2005){Br{\"u}ns}, {Kerp}, {Staveley-Smith},
  {Mebold}, {Putman}, {Haynes}, {Kalberla}, {Muller}, \&
  {Filipovic}}]{bruens05}
{Br{\"u}ns}, C., {Kerp}, J., {Staveley-Smith}, L., {Mebold}, U., {Putman},
  M.~E., {Haynes}, R.~F., {Kalberla}, P. M.~W., {Muller}, E., \& {Filipovic},
  M.~D. 2005, \aap, 432, 45

\bibitem[{{Choi} {et~al.}(2007){Choi}, {Weinberg}, \& {Katz}}]{choi07}
{Choi}, J.-H., {Weinberg}, M.~D., \& {Katz}, N. 2007, \mnras, 381, 987

\bibitem[{{Comeron} \& {Torra}(1992)}]{comeron92}
{Comeron}, F. \& {Torra}, J. 1992, \aap, 261, 94

\bibitem[{{Connors} {et~al.}(2006){Connors}, {Kawata}, \& {Gibson}}]{connors06}
{Connors}, T.~W., {Kawata}, D., \& {Gibson}, B.~K. 2006, \mnras, 371, 108

\bibitem[{{Kallivayalil} {et~al.}(2006{\natexlab{a}}){Kallivayalil}, {van der
  Marel}, \& {Alcock}}]{kallivayalil06b}
{Kallivayalil}, N., {van der Marel}, R.~P., \& {Alcock}, C. 2006{\natexlab{a}},
  \apj, 652, 1213

\bibitem[{{Kallivayalil} {et~al.}(2006{\natexlab{b}}){Kallivayalil}, {van der
  Marel}, {Alcock}, {Axelrod}, {Cook}, {Drake}, \& {Geha}}]{kallivayalil06a}
{Kallivayalil}, N., {van der Marel}, R.~P., {Alcock}, C., {Axelrod}, T.,
  {Cook}, K.~H., {Drake}, A.~J., \& {Geha}, M. 2006{\natexlab{b}}, \apj, 638,
  772

\bibitem[{{Knapp} {et~al.}(1978){Knapp}, {Tremaine}, \& {Gunn}}]{knapp78}
{Knapp}, G.~R., {Tremaine}, S.~D., \& {Gunn}, J.~E. 1978, \aj, 83, 1585

\bibitem[{{Konz} {et~al.}(2002){Konz}, {Br{\"u}ns}, \& {Birk}}]{konz02}
{Konz}, C., {Br{\"u}ns}, C., \& {Birk}, G.~T. 2002, \aap, 391, 713

\bibitem[{{Levine} {et~al.}(2006){Levine}, {Blitz}, \& {Heiles}}]{levine06b}
{Levine}, E.~S., {Blitz}, L., \& {Heiles}, C. 2006, Science, 312, 1773

\bibitem[{{Lu} {et~al.}(1998){Lu}, {Sargent}, {Savage}, {Wakker}, {Sembach}, \&
  {Oosterloo}}]{lu98}
{Lu}, L., {Sargent}, W. L.~W., {Savage}, B.~D., {Wakker}, B.~P., {Sembach},
  K.~R., \& {Oosterloo}, T.~A. 1998, \aj, 115, 162

\bibitem[{{Mastropietro} {et~al.}(2005){Mastropietro}, {Moore}, {Mayer},
  {Wadsley}, \& {Stadel}}]{mastropietro05}
{Mastropietro}, C., {Moore}, B., {Mayer}, L., {Wadsley}, J., \& {Stadel}, J.
  2005, \mnras, 363, 509

\bibitem[{{Mathewson} {et~al.}(1974){Mathewson}, {Cleary}, \&
  {Murray}}]{mathewson74}
{Mathewson}, D.~S., {Cleary}, M.~N., \& {Murray}, J.~D. 1974, \apj, 190, 291

\bibitem[{{Mathewson} {et~al.}(1979){Mathewson}, {Ford}, {Schwarz}, \&
  {Murray}}]{mathewson79}
{Mathewson}, D.~S., {Ford}, V.~L., {Schwarz}, M.~P., \& {Murray}, J.~D. 1979,
  in IAU Symposium, Vol.~84, The Large-Scale Characteristics of the Galaxy, ed.
  W.~B. {Burton}, 547--556

\bibitem[{{McClure-Griffiths} {et~al.}(2005){McClure-Griffiths}, {Dickey},
  {Gaensler}, {Green}, Haverkorn, \& Strasser}]{mcgriff05}
{McClure-Griffiths}, N.~M., {Dickey}, J.~M., {Gaensler}, B.~M., {Green}, A.~J.,
  Haverkorn, M., \& Strasser, S. 2005, \apjs, 158, 178

\bibitem[{{McClure-Griffiths} {et~al.}(2006){McClure-Griffiths}, Ford, Pisano,
  Gibson, Staveley-Smith, Calabretta, Kalberla, \& Dedes}]{mcgriff06a}
{McClure-Griffiths}, N.~M., Ford, A., Pisano, D.~J., Gibson, B.~K.,
  Staveley-Smith, L., Calabretta, M.~R., Kalberla, P. M.~W., \& Dedes, L. 2006,
  \apj, 638, 196

\bibitem[{{Nakanishi} \& {Sofue}(2003)}]{nakanishi03}
{Nakanishi}, H. \& {Sofue}, Y. 2003, \pasj, 55, 191

\bibitem[{{Nidever} {et~al.}(2007){Nidever}, {Majewski}, \& {Butler
  Burton}}]{nidever07}
{Nidever}, D.~L., {Majewski}, S.~R., \& {Butler Burton}, W. 2007, ArXiv
  e-prints, 706

\bibitem[{{Putman} {et~al.}(1998){Putman}, {Gibson}, {Staveley-Smith}, {Banks},
  {Barnes}, {Bhatal}, {Disney}, {Ekers}, {Freeman}, {Haynes}, {Henning},
  {Jerjen}, {Kilborn}, {Koribalski}, {Knezek}, {Malin}, {Mould}, {Oosterloo},
  {Price}, {Ryder}, {Sadler}, {Stewart}, {Stootman}, {Vaile}, {Webster}, \&
  {Wright}}]{putman98}
{Putman}, M.~E., {Gibson}, B.~K., {Staveley-Smith}, L., {Banks}, G., {Barnes},
  D.~G., {Bhatal}, R., {Disney}, M.~J., {Ekers}, R.~D., {Freeman}, K.~C.,
  {Haynes}, R.~F., {Henning}, P., {Jerjen}, H., {Kilborn}, V., {Koribalski},
  B., {Knezek}, P., {Malin}, D.~F., {Mould}, J.~R., {Oosterloo}, T., {Price},
  R.~M., {Ryder}, S.~D., {Sadler}, E.~M., {Stewart}, I., {Stootman}, F.,
  {Vaile}, R.~A., {Webster}, R.~L., \& {Wright}, A.~E. 1998, \nat, 394, 752

\bibitem[{{Quilis} \& {Moore}(2001)}]{quilis01}
{Quilis}, V. \& {Moore}, B. 2001, \apjl, 555, L95

\bibitem[{{Tenorio-Tagle} {et~al.}(1987){Tenorio-Tagle}, {Franco},
  {Bodenheimer}, \& {Rozyczka}}]{tenorio87}
{Tenorio-Tagle}, G., {Franco}, J., {Bodenheimer}, P., \& {Rozyczka}, M. 1987,
  \aap, 179, 219

\bibitem[{{van der Marel} {et~al.}(2002){van der Marel}, {Alves}, {Hardy}, \&
  {Suntzeff}}]{vandermarel02}
{van der Marel}, R.~P., {Alves}, D.~R., {Hardy}, E., \& {Suntzeff}, N.~B. 2002,
  \aj, 124, 2639

\bibitem[{{Wannier} \& {Wrixon}(1972)}]{wannier72}
{Wannier}, P. \& {Wrixon}, G.~T. 1972, \apjl, 173, L119

\bibitem[{{Yoshizawa} \& {Noguchi}(2003)}]{yoshizawa03}
{Yoshizawa}, A.~M. \& {Noguchi}, M. 2003, \mnras, 339, 1135

\end{thebibliography}

\clearpage

 \begin{figure}
 \centering
  \includegraphics[height=4in]{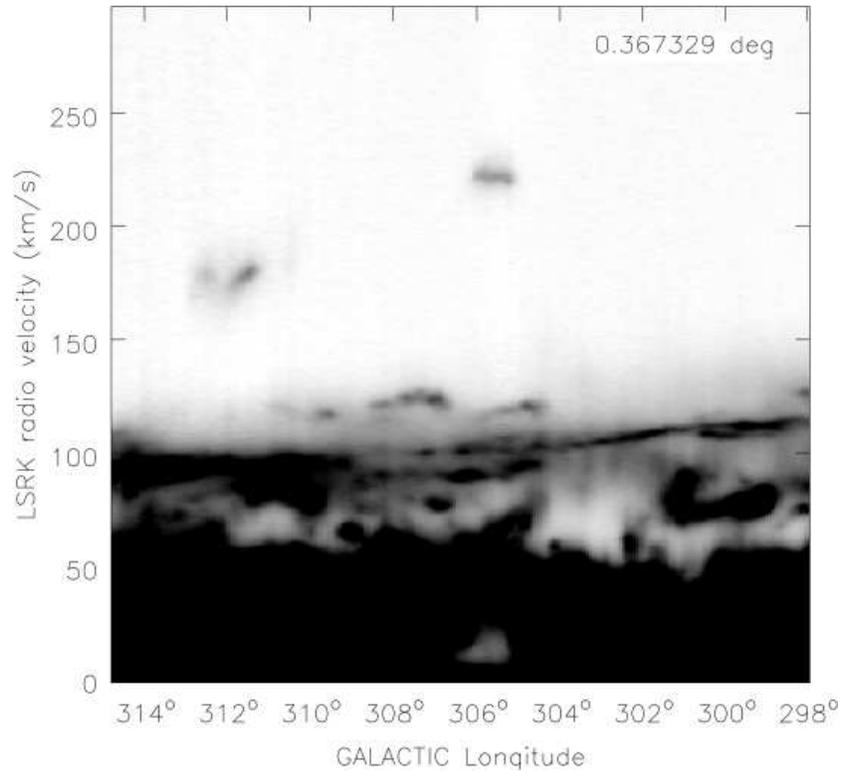}
  \caption[]{Longitude-velocity ($l$,$v$) display of the GASS \HI\
    data showing the ridge of disk \HI\ at $V_{LSR} \approx 120$ km
    s$^{-1}$ coincident with HVC 306-2+230. This longitude cut was
    taken through the head of the HVC at $b=0.37\arcdeg$.  The ridge is localized   between HVC 306-2+230 and another Leading Arm cloud, HVC
    312+1+180.  The
    greyscale is linear between -0.2 and 10 K.  
    \label{fig:vellong}}
 \end{figure}
 
\begin{figure}
\centering
\includegraphics[height=3.5in]{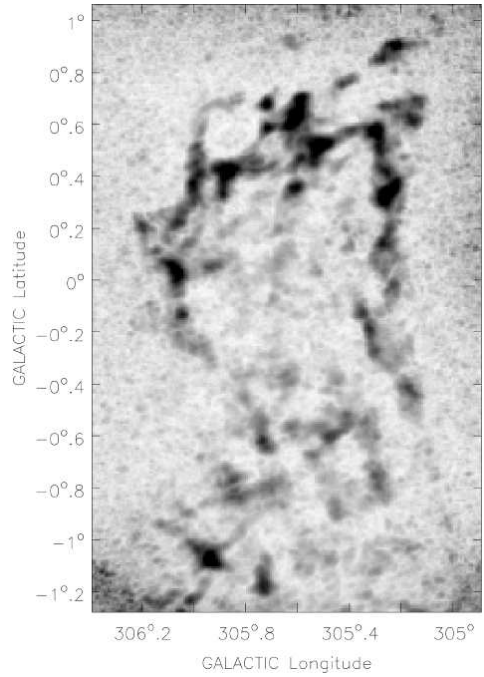}
\caption[]{High resolution \HI\ image of HVC 306-2+230.  The image is
  of \HI\ peak brightness temperature in the velocity range 175 - 300
  \kms, which covers the HVC only.  The greyscale is linear between 2 and 20 K.  The image has an angular
  resolution of 100\arcsec\ and a rms brightness temperature
  noise of $\sim 1.3$ K.  The cloud shows a very narrow ridge of
  emission around the head of the cloud, consistent with a
  bow-shock morphology.
  \label{fig:LA_sgps}}
\end{figure}

\begin{figure}
\centering
\includegraphics[width=4in]{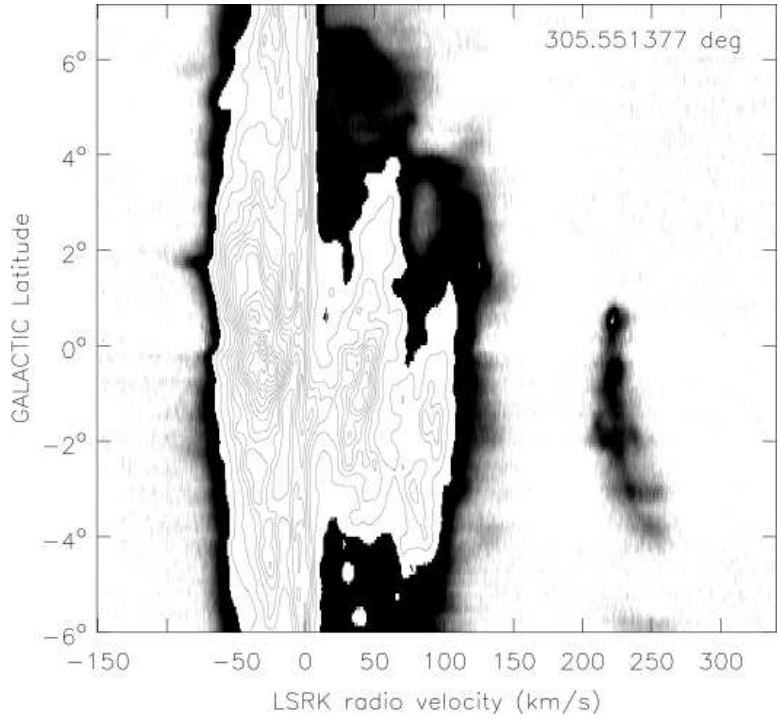}
\caption[]{Latitude-velocity image of \HI\ emission at $l=306\arcdeg$
  showing the velocity of HVC 306-2+230 changing from 223 \kms\ at
  $b=-1.5\arcdeg$ to 252 \kms\ at $b=-4.4\arcdeg$. The
  greyscale has a power law scaling of $-0.5$ and goes between $0-2$
  K.  \HI\ emission brighter than 5 K is displayed only in contours.  
  We interpret the velocity gradient in the HVC as arising from drag caused by 
  its encounter with the Galactic disk.
  \label{fig:LA_bv}}
\end{figure}

\end{document}